\documentstyle[11pt,aaspp]{article}

\tighten

\newcommand{\gtsim}{\ {\raise-0.5ex\hbox{$\buildrel>\over\sim$}}\ }
\newcommand{\ltsim}{\ {\raise-0.5ex\hbox{$\buildrel<\over\sim$}}\ }
\def\Msun{\hbox{$\thinspace M_{\odot}$}}
\def\Lsun{\hbox{$\thinspace L_{\odot}$}}
\def\Zsun{\hbox{$\thinspace Z_{\odot}$}}
\def\simlt{\lower.5ex\hbox{$\; \buildrel < \over \sim \;$}}
\def\simgt{\lower.5ex\hbox{$\; \buildrel > \over \sim \;$}}

\journalid{}{}
\articleid{11}{14}  

\begin{document}

\title{On the Effects of Bursts of Massive Star Formation During the 
Evolution of Elliptical Galaxies}
 
\author{Stephen E. Zepf\altaffilmark{1}}
\affil{Department of Astronomy, University of California,
    Berkeley, CA 94720 \\ e-mail: zepf@astron.berkeley.edu}

\author{Joseph Silk}
\affil{Departments of Physics and Astronomy and Center for Particle 
	Astrophysics, \\ University of California, Berkeley, CA 94720 
	\\ e-mail: silk@pac2.berkeley.edu}

\vskip 24pt

\centerline {To be published in {\it The Astrophysical Journal}}

\vskip 24pt

\altaffiltext{1}{Hubble Fellow}

\begin{abstract}

	We consider the hypothesis that the formation of elliptical
galaxies includes a phase in which star formation is mostly restricted
to massive stars, with the bias towards high mass stars increasing
with elliptical galaxy mass. The mass fraction of stars in this
top-heavy mode of star formation is constrained by requiring the resulting 
stellar remnants to account for the observed increase in the mass-to-light 
ratio of ellipticals with increasing galaxy mass. We then consider
the implications of this population of massive stars for the
intracluster medium and the extragalactic background at various
wavelengths. The mass and abundance ratios of metals produced by 
our proposed population of massive stars are consistent with the 
observations of the mass and abundance ratios of metals in the 
hot gas of galaxy clusters for most of the standard range of IMF 
slopes and SN II yields. The predicted energy density produced by 
this stellar population approaches current limits on the extragalactic 
background at both optical wavelengths, into which the ultraviolet
radiation of the massive stars is likely to be redshifted, and 
far-infared wavelengths, at which starlight reprocessed by dust 
associated with the starburst will be observed. In either case, 
the background is predicted to be significantly clustered since 
massive ellipticals are clustered.

\end{abstract}


\newpage

\section{Introduction}

	One of the remarkable features of elliptical galaxies is
the very narrow dispersion in the relationship between effective radius, 
surface brightness within that radius, and the central velocity dispersion
(Dressler {\it et al.} 1987, Djorgovski \& Davis 1987). This ``Fundamental
Plane'' of elliptical galaxies can be written as
$$ R_{e} \propto \sigma^{A} \Sigma_{e}^{B}, $$ 
where, in the $B$ band, the exponents A and B are 1.3 and $-0.8$ 
respectively with about a $10\%$ uncertainty.
The deviation of this relationship from a purely virial one leads
to the scaling between mass-to-light ratio and mass,
$(M/L)_B \propto M^{0.2}$ (Faber {\it et al.} 1987).
More recently, near-infrared studies have shown that the Fundamental
Plane in the $K$ band yields $(M/L)_K \propto M^{0.16}$
(Pahre, Djorgovski, \& de Carvalho 1995, Mobasher {\it et al.} 1996).

	The question arises as to the nature of the physical mechanism driving 
the scaling between $M/L$ and $M$. One potential candidate is metallicity, 
since metallicity is believed to increase with elliptical galaxy 
mass/luminosity ratio, and increasing metallicity leads to increasing
$M/L$ in optical bandpasses. However, the effect of metallicity 
variations on the observed $M/L$ trend is strongly limited by
the observation that the scaling between $M/L$ and $M$ extends
into the K-band, because according to stellar models $(M/L)_K$ 
is insensitive to metallicity for old stellar populations.
This result is in agreement with earlier work which showed that the
change in metallicity inferred from the slope of the color-magnitude
relationship leads to  $M/L$ changes that are  much less than observed
(Dressler {\it et al.} 1987). As a specific example, we estimate
the increase of metallicity with luminosity from the observed
$K, V-K$ relationship and the models of Charlot \& Bruzual (1996).
These models suggest metallicity variations affect the the exponent 
of the scaling between $M/L$ and $M$ by less than several-hundredths 
in the $K$ band.

 	A second possibility is that the ages of ellipticals may vary 
systematically with galaxy mass/luminosity. In order to obtain the 
correct sense of the $M/L$ trend, low luminosity ellipticals must be 
much younger than high luminosity ellipticals. We find that variations 
in the mean stellar age of ellipticals such that the V-K differences 
are matched can account for about $1/2$ of the observed (M/L) scaling. 
The resulting ages range from about 6.5 Gyr for $\sim 0.1 L_{*}$ 
ellipticals to about 17 Gyr for $\sim 3 L_{*}$ ellipticals.
Such large age differences seem hard to reconcile with the small
rms scatter in $(U-V)$ colors for the same ellipticals ({\it e.g.} less than 
$0.^m 04$ in Coma, according to Bower, Lucey \&  Ellis 1992). Unless the epoch
at which star formation terminated was remarkably well coordinated,
age variations at a level significant for the $M/L$ scaling
would appear to be ruled out. A finely adjusted combination of
age and metallicity might somewhat weaken this constraint.

	Another alternative is that the structure of elliptical galaxies 
may vary with luminosity. This can lead to inferred mass-to-light variations 
when the mass is derived from the central velocity dispersion and
elliptical galaxies are assumed to be homologous. One possibility
is that the change from rotational support to velocity anisotropy with 
elliptical galaxy luminosity may affect the inferred mass-to-light ratios 
({\it e.g.} Prugniel \& Simien 1994), although other studies have found 
that the effect on the Fundamental Plane is small 
({\it e.g.} Djorgovski \& Santiago 1993). Alternatively, the density profiles 
of elliptical galaxies may vary with luminosity ({\it e.g.} Caon, Capaccioli, 
\& D'Onofrio 1993), possibly leading to some effect on the scaling 
of $M/L$ with $M$ (see also Capelato, de Carvalho, \& Carlberg 1995).
A variation on this overall theme which appeals to changing the radial 
distribution of cold dark matter halos as a function of elliptical galaxy 
luminosity was proposed by Guzm\'an, Bower, \& Lucey (1993). 

	Systematic variations of the stellar initial mass function (IMF)
can also produce a trend of $M/L$ with $M$ (e.g. Djorgovski 1988).
Most earlier work concentrated
on increasing $M/L$ for more massive/luminous ellipticals by tying up more 
mass in stars well below the main-sequence turn-off through a steeper IMF
(Djorgovski \& Santiago 1993, Renzini \& Ciotti 1993). However,
such an IMF leads to stellar abundance trends opposite
to those observed, which indicate preferential enrichment from
the products of massive stars in the most massive ellipticals
(e.g. Worthey, Faber, \& Gonzalez 1992, Matteucci 1994).
An alternative, first suggested in this context by Larson (1986), 
is a bimodal IMF, with stellar remnants from an earlier generation
of massive stars as an explanation for the increase of $M/L$
with $M$.

	 In this paper, motivated by other considerations, we further
develop the idea that a bimodal IMF, with stellar remnants contributing
significantly to the mass, is responsible for the increase of both
$M/L$ and metallicity with increasing galaxy mass. {\it Our hypothesis 
is that the increase of $M/L$ with $M$ derived from the analysis of 
the Fundamental Plane is the result of a population of remnants of 
massive stars which makes up a systematically larger fraction of the 
mass of the galaxy with increasing galaxy mass, and that the associated
supernova ejecta are responsible for explaining an outstanding problem
that is closely associated with the presence of early-type galaxies,
namely the enrichment of the intracluster medium.} 
We associate the episode of formation of the massive star population 
with major starbursts, most likely induced as the result of early galaxy 
mergers which may play an important role in elliptical galaxy formation. 

There are both observational and theoretical
grounds for the proposal that the initial mass function (IMF) in 
starbursts is strongly biased towards massive stars. 
Observationally, several studies of the central regions of starbursts 
have suggested that the IMF is strongly biased towards high-mass stars 
({\it e.g.} Rieke {\it et al.} 1980, Doane \& Mathews 1993,
Rieke {\it et al.} 1993, Doyon, Joseph, \& Wright 1994). 
From the theoretical 
standpoint, it has been argued that such an IMF biased towards massive
stars is expected in starbursts, especially when a merger is involved.
In such a situation, the turbulence in the interstellar 
medium is much higher than in the Galactic regions for which standard 
IMFs have been determined, and may play a central role in modifying 
the IMF (Silk 1995). This effect is likely to be strongest in the 
starbursts which are the progenitors of the most massive ellipticals, 
and which have the deepest potential wells. 

There is also a variety of circumstantial evidence suggesting that
physical conditions during periods of star formation in elliptical
galaxies were dependent on galaxy mass. These include observations 
of nearby ellipticals which show that
[Mg/Fe] increases above the solar value with increasing
galaxy luminosity (Worthey {\it et al.} 1992, 
Davies, Sadler, \& Peletier 1993), 
and of distant galaxies with unusually red colors or strong
4000 \AA\ breaks (Charlot {\it et al.} 1994). These results are suggestive 
of an IMF that once was weighted towards massive stars, perhaps during 
the early merger that was the hallmark of a forming elliptical. Also, the 
specific frequency of globular clusters increases with elliptical 
galaxy luminosity (Djorgovski \& Santiago 1992; Zepf, Geisler, \& Ashman 1994),
and may be indicative of a systematic change in the way stars form 
with increasing potential well depth. Formation of globular clusters does
indeed appear to be a signature of major mergers 
({\it e.g.} Whitmore \& Schweizer 1995, Zepf {\it et al.} 1996).

	A top-heavy IMF during the initial stage of elliptical galaxy
formation has also been proposed as a way to account for the large
iron masses in galaxy clusters 
({\it e.g.} Elbaz, Arnaud, \& Vangioni-Flam 1995).	
Recent observations of the elemental abundances of the intracluster
gas point to type II SN an the source of enrichment, providing
strong support for models with a top-heavy IMF during the formation
of elliptical galaxies (Loewenstein \& Mushotzky 1996). We will
return to the issue of the enrichment of the ICM in Section 3.

	The primary goal of this paper is to test the hypothesis
that the trend of higher $M/L$ with increasing $L$ is the result of
compact stellar remnants from a past burst of formation of high-mass stars
in massive ellipticals. Our approach is to fix the remnant population
to match the Fundamental Plane results, and then consider the 
observational implications of the massive star population required
to produce these remnants. We note that our focus is on the slope of
the Fundamental Plane, and not on its narrowness. A study of the 
small scatter about the fundamental plane, which implies a high 
degree of synchronization at any given elliptical galaxy mass/luminosity 
(e.g. Renzini \& Ciotti 1993, Djorgovski \& Santiago 1993), 
is beyond the scope of this paper.

Fortunately, our proposed top-heavy mode of star formation during 
the evolution of giant elliptical galaxies produces 
a number of phenomena accessible to observations.
One of the most direct tests of the model is the metal yields of the 
massive star population. We particularly concentrate on the predicted 
metallicity and abundances of the intracluster medium, for which new
data exist. A second test of the model is its contribution
to the extragalactic background light at a variety of wavelengths. 
We find that the model is consistent with the observations, and that
the effects of the proposed stellar population
may have been observed. We then describe further observational tests
of the initial hypothesis. The paper is laid out such that the basic
properties of the proposed population are calculated in $\S$ 2, comparisons 
to current observations are given in $\S$ 3, and $\S$ 4 contains 
the discussion of future observational tests of our hypothesis.

\section {Analysis}

	Our procedure is to first calculate the mass in remnants required
at each elliptical galaxy luminosity in order to explain the observed
$M/L$ trend. We then integrate over the luminosity function for early-type 
galaxies to obtain the global mass density of these remnants. Finally, 
we use models of stellar populations to predict the metal yields produced 
by the massive stars which result in the calculated mass in remnants. 

	In order to determine the fraction of galaxy mass in remnants
of massive stars as a function of luminosity, we adopt a fiducial 
elliptical galaxy luminosity for which all of the stars form with 
a standard IMF. One natural choice for this luminosity is the point 
at which the surface brightness--absolute magnitude relationship 
changes slope, indicating a shift from 
the population of normal, giant ellipticals to the separate population 
of dwarf ellipticals (Kormendy \& Djorgovski 1989 and references therein).
This shift takes place at roughly $M_{B} = -17 + 5{\rm log}h$,
or about 0.1 $L_{*}$.
We take this to be the luminosity of an elliptical with a ``normal'' 
stellar population, and any increase in M/L for more luminous ellipticals 
to be the result of stellar remnants. The choice of 0.1 $L_{*}$
is not a firm one, both because the luminosity of the shift from
giant to dwarf ellipticals is not tightly constrained 
({\it e.g.} Sandage \& Perlmuter 1991) and because 
other effects may play a role in the relationship between $M/L$ and $M$. 
For example, if we used the shift from dynamical support primarily
through rotation as the defining point for an elliptical galaxy
with a ``normal'' stellar population, the luminosity would be a factor
of a few higher than 0.1 $L_{*}$, albeit with large uncertainties.
Similarly, the elliptical galaxy sequence may extend to luminosities
fainter than 0.1 $L_{*}$. Therefore, we also consider 
the effect of defining the luminosity at which an elliptical has a ``normal''
stellar population as 0.03 $L_{*}$ or 0.3 $L_{*}$. We will refer to a
fiducial luminosity of 0.1 $L_{*}$ as case one, and these latter two
luminosities as cases two and three respectively.


	For the (M/L)$_B$ of early-type galaxies at $L_{*}$, we adopt $13h$.
This is an average of the observed values, which range from $11.9h$ 
(van der Marel 1991), $12.7h$ (Bender, Burstein, \& Faber 1992), and 
$14.5h$ (Lauer 1985), where these have all been converted to values
at $L_{*}$. With $(M/L) \propto L^{0.2}$, these values give 
(M/L)$_B$ = 8.2h at $0.1 L_{*}$, and (M/L)$_B$ = 15.6h at $2.5 L_{*}$, 
which is roughly the luminosity of M87. Thus, for case 1 in which
ellipticals at 0.1 $L_{*}$ have only ``normal'' stellar populations,
about one-third of the mass in ellipticals at $L_{*}$ is in remnants 
of massive stars, and this fraction is almost one-half for ellipticals 
like M87. For cases 2 and 3, in which 0.03 $L_{*}$ and 0.3 $L_{*}$
galaxies are ``normal'', the fraction of mass in remnants of massive 
stars is about 60\% and 33\% respectively for ellipticals like M87.
Figure 1 shows graphically how changes in the lower luminosity limit
lead to changes in the mass fraction of remnants as a function of
luminosity.



	For the luminosity function of early-type galaxies, we adopt
the usual Schechter form (Schechter 1976) with 
$L_{*} = 1.0 \times 10^{10} h^{-2}$ \Lsun\ (in the $B$ band), 
$\alpha = -0.3$, and $\phi_{*} =1.0 \times 10^{-2} h^{3}$ Mpc$^{-3}$.
This luminosity function is based primarily on the recent results
of the Las Campanas Redshift Survey (LCRS) for galaxies without 
emission lines (Lin {\it et al.} 1995). Our adopted luminosity function 
is also similar to that found in the APM-Stromlo survey for galaxies
classified morphologically as early-type (Loveday {\it et al.} 1992). 
The differences between the two surveys are relatively minor for 
our purposes, with Loveday {\it et al.} finding a slightly more positive 
$\alpha$, brighter $L_{*}$, and smaller $\phi_{*}$ such that the overall 
luminosity density is very similar. A somewhat different luminosity 
function for early-type galaxies was derived from the shallower CfA redshift 
survey by Marzke {\it et al.} (1994). They found $\alpha = -0.9$, and 
a density normalization about a factor of two higher than the other 
two surveys. We will consider the effects of adopting these latter 
values for the luminosity function, although the significantly greater 
volume of the former two surveys suggests that they are more likely to 
adequately sample the true luminosity function of ellipticals at
the present epoch.


	To calculate the total mass fraction of the population of the
remnants of massive stars, we convolve the mass estimates with the
luminosity function for early-type galaxies. The resulting mass
function of remnants is shown in Figure 2. This figure also shows
the effect of adopting cases 2 and 3 for the cutoff luminosity
for elliptical galaxies. For all cases, the dominant contributors
to the remnant mass function are elliptical galaxies with
luminosities of about $2 L_{*}$. 

	The total mass density of the remnant population is the
integral of the mass function shown in Figure 2. For the fiducial case,
this density of remnants from the episode of massive star formation 
which produced the $M/L$ vs. $M$ dependence of elliptical galaxies is
$3.7 \times 10^{8} h^{2}$ \Msun\ Mpc$^{-3}$. If we change the
cutoff luminosity for elliptical galaxies, the densities are
170\% and 40\% of the fiducial density, for cases 2 and 3
respectively. Also, if we change the luminosity function to
one like that of Marzke {\it et al.} (1994), we find that the remnant
density is very similar to that given by the fiducial values. 
The reason for this result is that the remnant density is decreased
by the more negative $\alpha$, but increased by a similar amount
by the higher normalization. For the remainder of the paper, 
we adopt a global remnant mass density of 
$3.7 \times 10^{8} h^{2}$ \Msun\ Mpc$^{-3}$, and assume
this value is uncertain by about a factor of two.


	The next step is to estimate the yield of metals from the burst 
of massive stars which produced this density of remnants. This requires 
the IMF of the starburst, and the metal yield and remnant mass as a 
function of the stellar mass. For the IMF of the starburst, we rely 
on observations of starburst galaxies at the current epoch, which suggest 
that the IMF in these regions may have a lower cutoff at about 
3 \Msun\ ({\it e.g.} Rieke {\it et al.} 1993). 
On this basis, we adopt an IMF with a Salpeter slope ($x=1.35$), 
$m_{L} = 3$\Msun\, and $m_{U} = 60$\Msun . We also consider how variations 
in the IMF affect the conclusions. 
In Figure 3, we plot these IMFs and a Miller-Scalo (1986) IMF representing
the normal elliptical galaxy population for comparison.
We note that the minimum requirement 
for the IMF of the starburst is that it is restricted to masses greater 
than about 1\Msun\ so that these stars are not luminous today. 

	For the yield of metals from these massive stars, we use
the recent Woosley \& Weaver (1995) determinations for SN II
of various masses. For simplicity, we take the average of their
different models for stars of $M \ge 30$ \Msun . The differences
between different physical models for these very massive stars
produce different net yields of about $20\%$ for the IMFs we consider.
The models of Thielemann, Nomoto, \& Hashimoto (1995) also produce net
yields with differences of this order or somewhat greater. The
Thielemann {\it et al.} models produce more Mg and less Fe for higher
mass stars than the Woosley \& Weaver models.

We track the elements Fe, Mg, and O, as these are the most 
readily observed in the stars of elliptical galaxies and the hot 
gas in clusters of galaxies. For O, we also consider the yields 
from stars with $M < 10$\Msun\, using the yields of Renzini \& Violi (1981). 
The final ingredient required are the final remnant masses produced,
for which we use the values of Prantzos {\it et al.} (1993). For massive
stars, these are similar to those expected from the Woosley \& Weaver
models.

	The result for our fiducial IMF is that, per solar mass of
remnants, a total mass in metals of about  $4.2 \times 10^{-1}$ \Msun\ 
is produced, with 
specific element production of $1.3 \times 10^{-2}$ \Msun\ of Fe, 
$9.9 \times 10^{-3}$ \Msun\ of Mg, and $2.2 \times 10^{-1}$ \Msun\ of O. 
These values can then be multiplied by the mass density of remnants 
for cases 1 and 2 derived above to obtain the global mass density of 
various elements. Flatter IMFs produce more metals with increases in 
[O/Fe] and [Mg/Fe]; steeper IMFs produce fewer metals with a somewhat 
lower [Mg/Fe] and [O/Fe] ratios. Raising the upper mass limit increases 
the yields of O and Mg slightly, and lowering it decreases them slightly. 
Lowering the lower mass limit decreases the yields of O and Mg and also 
Fe slightly, whereas raising it primarily increases the overall yield, 
as long as $M_{low} < M_{SN II}$.
Use of the Miller-Scalo IMF, peaking near $0.3$ \Msun\ 
but with a flatter IMF slope above 1 \Msun\, 
would provide an alternative to a truncated IMF for the starburst, 
but at the price of enhancing 
the SN I/SN II rate and thereby reducing the yield for a given $M/L$ value.



\section {Observational Implications}

\subsection {Metal Production}

	Accounting for the metals produced by this burst of high-mass 
star formation is a critical test of the model. This test is straightforward 
in the sense that the model makes specific predictions about the masses
of various metals and which galaxies produce them. Moreover, it is 
expected that a starburst will drive a strong galactic wind, particularly 
when predominantly high mass stars are formed ({\it e.g.} Doane \& Mathews 1993). 
This expectation is supported by observations of ``superwinds'' driven 
by starbursts ({\it e.g.} Heckman, Armus, \& Miley 1990).  To first order, 
most of the metals produced will end up in the intergalactic medium
(IGM), or in the case of cluster galaxies, the intracluster medium (ICM). 
The real situation is undoubtedly more complex, involving the interplay 
of structure evolution, gravitational interactions, star formation, 
hydrodynamics, and chemical evolution. However, a useful starting point 
is the comparison between the metals produced in the proposed burst, 
and those observed in the IGM or ICM.

	The comparison with the ICM is particularly interesting, because 
abundances can be derived from X-ray observations of hot gas in clusters, 
and because these regions are rich in early-type galaxies. As a specific 
example, we consider the well-studied Coma Cluster. For Coma, the 
overall luminosity is $L_{B} = 1.9 \times 10^{12}\ h^{-2}$ \Lsun\ 
({\it e.g.} Godwin \& Peach 1977, Kent \& Gunn 1982). Taking this luminosity, 
accounting for the contribution of galaxies other than early-types,
and plugging in to our previous values for the mass of remnants and 
Fe at a given number density of early-type galaxies, we find a mass 
in remnants of $M_{rem} = 5.9 \times10^{12}\ h^{-1}$ \Msun\, and a 
Fe mass of $M_{Fe} = 7.7 \times10^{10}\ h^{-1}$ \Msun .
For the mass in Fe inside an Abell radius, we convolve the
White {\it et al.} (1993) estimate of the gas mass 
$M_{gas} = 5.45 \times 10^{13}\ h^{-5/2}$ \Msun\  in Coma with its
Fe abundance, which is about 0.35 solar (Ohashi 1995)
and find $M_{Fe} = 2.3 \times 10^{10}\ h^{-5/2}$ \Msun .
Thus given the assumption that all of the metals produced by the
burst of massive stars are ejected into the ICM, our fiducial case predicts
an Fe mass in Coma which is $3.4 h^{3/2}$ greater than that observed.

	This comparison is shown graphically in Figure 4, in which
we plot the predicted $M_{Fe}$ as a function of the luminosity at
which an elliptical galaxy is fixed to have a ``normal'' stellar
population. The plot demonstrates that the model with the fiducial values
is a reasonable fit to the data if $h$ is close to 0.5, and requires 
a somewhat brighter cutoff luminosity for ellipticals if $h$ is closer 
to 1.0. An alternative approach is to follow Arnaud {\it et al.} (1992)
and Renzini {\it et al.} (1993) and adopt a typical ``iron mass to light 
ratio (IMLR).'' These authors show that observed IMLR values are
roughly $M_{Fe}/L_{V}^{E/S0} = 0.02 h^{-1/2}$. For our fiducial
values, we find that this ratio is 0.039 h.

	We note that these calculations are based on the assumption
that all of the metals from the burst of massive stars are ejected into
the ICM. This probably represents an upper limit to the metals contributed 
to the ICM, since some of the metals are likely to be locked up in 
generations of ``normal'' star formation which occur at the same time
or after the phase of massive star formation. The suprasolar values 
of [Mg/Fe] observed in massive ellipticals are indicative of enrichment
from an episode of massive star formation ({\it e.g.} Worthey {\it et al.} 1992,
Matteucci 1994). We can roughly estimate the amount of mass in metals locked
up in this way by requiring that all of the ``excess'' Mg observed 
in the stars of ellipticals originated in the burst of massive stars. 
This represents about half of the total metal production of the massive 
star formation phase. Thus the above predictions  for the model production 
of Fe are reduced by about a factor of two.

	We find that the best prediction of our fiducial model is
that it produces an Fe mass of 1-2 $h^{3/2}$ times the observed Fe mass
in clusters. This agreement is very promising, and well within the 
uncertainties of the calculation. These uncertainties include the 
luminosity function of early-type galaxies, the slope and limits 
of the top-heavy IMF, the yields from SN II, the adopted ratio  
$M_{Fe}/L_{V}^{E/S0}$, which is higher given a radially increasing 
gas to luminous mass ratio and lower with a declining radial gradient 
in metallicity, as well as the luminosity at which elliptical galaxies 
have ``normal'' stellar populations. We conclude that the fiducial
model successfully predicts the Fe mass in galaxy clusters, and look
for other observational tests.

	The abundance ratios of various elements provide a powerful
test of theories for the enrichment of the ICM. The prediction
of our fiducial model is fairly simple; the abundance ratios should
be like the yields from type II SN. Recent ASCA data for four clusters
find almost exactly such abundance ratios (Mushotzky {\it et al.} 1996). 
Thus the X-ray determined abundance ratios favor SN II
as the source of the enrichment of the ICM, independently of any specific
model (Loewenstein \& Mushotzky 1996). 
The strength of our model is that it naturally accounts for both the 
total Fe mass and the abundance ratios in the ICM, given only the 
constraint that the modified IMF accounts for the $M/L$ trend of 
elliptical galaxies in terms of compact remnants of massive stars.

	Massive elliptical galaxies are not located solely in rich
clusters, so the IGM will also be enriched by massive ellipticals
in poor clusters and groups. Most of the ejecta are unlikely to be 
retained by galaxy groups, and will become part of the diffuse IGM.
The observational constraints on the metal content of the IGM are 
few. However, we note that the global enrichment is expected to be
substantial. Given our fiducial model and the adopted luminosity function 
for early-type galaxies, the mass density of metals is 
$1.6 \times 10^{8}\ h^{2}$ \Msun\ Mpc$^{-3}$. About half of this mass
is likely to have been ejected from galaxies, leading to an
inferred enrichment of the IGM, 
$Z_{IGM}/\Zsun\approx  0.01 \Omega_{IGM}^{-1}.$

\subsection{Extragalactic Background Light}

	We have shown that a starburst composed of high mass stars
constrained to account for the change in M/L with L observed in
elliptical galaxies produces metal abundances in general
agreement with that observed in the ICM. Another implication of
such a burst is its effect on the extragalactic background light, 
as the energy density produced by the large numbers of high mass stars
may be significant. The energy density of a population
of stars can be written as a function of the metal density it
produces, $$\nu i_{\nu} = \epsilon c^{2} (\rho Z) c/4\pi (1+z_{f})^{-1}, $$
where $z_{f}$ is the redshift at which the starburst occurs, and
the value of $\epsilon$ is about 0.004 for stars in the mass range
of interest (Bond, Arnett, \& Carr 1984).
This calculation is straightforward for our model, because we
have previously found the mass density of metals to be
$1.6 \times 10^{8}\ h^{2}$ \Msun\ Mpc$^{-3}$, or 
$(\rho Z) = 1.1 \times 10^{-32}\ h^{2}\ {\rm g}\ {\rm cm}^{-3}$
in more conventional units for this type of work.
The resulting energy density is then 
$\nu i_{\nu} = 86~ (1+z_{f})^{-1}~ h^{2}~ {\rm nW}~ {\rm sr}^{-1}$.
For canonical values of $z_{f}$ of about 3, 
this is $22 h^{2}\ {\rm nW}~ {\rm sr}^{-1}$. 
	
Most of the energy of massive stars is radiated in the
ultraviolet. If these massive starbursts typically occur at moderately
high redshift, and there is little dust, this energy density will
be seen as a diffuse optical extragalactic background. The observational
limit on the background at $4000$~\AA\ is $20\ \rm nW\ {\rm sr}^{-1} $
(Mattila 1990). As shown in Figure 5, the prediction of the fiducial 
model is somewhat lower than these upper limits for typical values of $h$.
There is also reason to believe that the burst of massive star
formation may be dust-shrouded. There are close analogies between 
this phase of elliptical galaxy formation and ultraluminous starbursts: 
the rate and efficiency of star formation are comparable, and the
luminosity profiles of merger products, in the post-starburst phase, 
appear to be evolving towards $r^{1/4}$ profiles (e.g. Schweizer 1990,
Wright et al. 1990). 
Moreover, while the mild evolution 
observed for ellipticals in clusters at $z\sim 0.4$ argues for the 
formation of the bulk of the stars at $z$ \simgt 2, ({\it e.g.} Ellis 1996, 
Bender {\it et al.} 1995), unsuccessful searches for luminous forming 
galaxies to $z\sim 5$ (Pahre \& Djorgovski 1995) may be accounted 
for if protoellipticals are dusty.


	If the burst of massive star formation is shrouded in dust,
the ultraviolet radiation from massive stars will be reprocessed
into the far-infrared. The radiation density of local starbursts
peaks at about $60 \mu$, which for typical formation
redshifts, would be detected at present at roughly $240 \mu$.
The DIRBE instrument on the COBE satellite is sensitive to diffuse
backgrounds at these wavelengths. Hauser (1995) gives an upper limit
at $240 \mu$ of $\nu i_{\nu} \le 20 \rm nW\,sr^{-1}$, and 
Puget {\it et al.} (1995) argue that a diffuse background is detected
in these data at the level of
$\nu i_{\nu} \approx ~3-10 \rm nW\,sr^{-1}$
at similar wavelengths. 
The best constraints on an extragalactic background at longer wavelengths
come from the stringent limits on deviations from a blackbody 
in the FIRAS spectrum of the CMB (Mather {\it et al.} 1994).
Blain and Longair (1993) consider a number of models, including some whose
evolution and metal production are similar to ours, and find 
consistency with the FIRAS data. However, the results 
are dependent on the temperature and other physical properties of the dust,
and low dust temperatures can raise the predictions above the FIRAS limit.

	By analogy with starbursts, it is more likely that the correct 
model is a mixture of the two cases described above, rather than one
in which either all or none of the ultraviolet light is absorbed by dust 
and reradiated in the far-infrared. The same class of objects would 
then be responsible for most of the diffuse background at optical and 
far-infrared wavelengths.
A prediction of this scenario is that the cosmic backgrounds
in the optical and the far-infrared are correlated with one another.
Moreover the spatial structure of both of these backgrounds should reflect
the enhanced correlations of ellipticals relative to spirals.

\section{Discussion}

	We have explored the consequences of the hypothesis that remnants 
of massive stars make up a systematically larger fraction of the masses 
of elliptical galaxies with increasing galaxy mass such that they account 
for the scaling $(M/L)_{B} \propto M^{1/6}$. 
We find that this hypothesis is in good agreement with 
observations of metals in clusters and extragalactic background light. 
In fact, the observations seem to lead in the direction of models such 
as that explored here. The observational confirmation that the elemental 
abundances in the hot cluster gas resemble those of SN II is 
indicative of an early population of massive stars as the source of these 
metals (Loewenstein \& Mushotzky 1996). 
The suprasolar [Mg/Fe] ratios of massive ellipticals make them attractive 
candidates for the location of the stellar population that incorporates 
some of the ejecta from these massive stars. The increasing 
mass-to-light ratio with elliptical galaxy mass provides a natural 
place for the remnants of the massive star population which produced 
the SN II. There is no unique inference of an IMF, since our constraint 
is an integral one. However we find it remarkable that to within the 
uncertainties in the comparisons of about a factor of two, a consistent 
model can be constructed accounting for all of these observations.

We note that the supernova rate per unit luminosity required 
by our model of ICM enrichment is enhanced relative to conventional 
Galactic models by a factor of $\sim 5$, the ratio of $M_{Fe}/M_\ast$ 
in clusters to its value in the Milky Way, for which the ratio of 
SN Ia to SN II that reproduces the observed abundances 
is approximately 0.15 (Tsujimoto {\it et al.} 1995).
Since SN Ia leave no compact remnants, our model and the observed 
abundance ratios require the ICM Fe to be produced almost exclusively 
by SN II. Hence the rate of SN II in protoellipticals is enhanced by 
a factor of about 6 relative to 
the SN II rate requirements of the standard chemical evolution model for
the Galaxy. What we have not addressed here in any detail are the means 
by which the enriched debris is ejected into the ICM and IGM. 
A galactic wind is likely, and given the high supernova rate may well 
be inevitable even from massive galaxies.
Our model predicts about 0.2 SN II per \Msun\ of remnants.
We will discuss elsewhere the implications of such a high supernova rate
for the interstellar medium in the starburst.
This may lead to important consequences for such 
supernova byproducts as cosmic rays
and gamma rays. Although the details are sensitive to the low energy 
cosmic ray spectrum, it would be of interest to compute the effects 
of spallation and LiBeB production as well as of gamma rays from 
nuclear excitations and $\pi^0$ decays arising from cosmic ray 
interactions with the dense protogalactic interstellar gas.

	The formation of the large number of massive stars
indicated by our model may have a significant effect on the
extragalactic background light. We find that our fiducial
model produces an extragalactic background just below the
current observational limits. Studies of the extragalactic background
at various wavelengths are probably the most promising way to further
constrain our hypothesis. One aspect of our model is that the background
is expected to be clustered, since most of it is produced by bright
elliptical galaxies, which are known to be significantly clustered.
This clustering may help distinguish our model from other models,
such as a simple luminosity evolution in which all sources become 
brighter in the past or density evolution in which the number of
sources at all luminosities increases 
(e.g. Lonsdale {\it et al.} 1990, Saunders {\it et al.} 1990).
However ``backwards evolution'' models like these have an ad hoc 
normalization which is typically set by requiring that the models 
not overproduce the cosmic infrared background. Alternatively, models
like those of Franceschini {\it et al.} (1994) are based on the local
luminosity function and a prescription
for the star formation history of different galaxy types.
The predictions of our models for observations at far-infrared and sub-mm
wavelengths are equivalent to a luminosity-dependent density evolution,
with the evolution concentrated on the brightest and most clustered objects,
and with the normalization fixed by the observed metallicity production.

The bursts of massive star formation may lead to bright sources
which are individually detectable, particularly if the
burst occurs on a fairly short timescale. 
Searches for these discrete sources in the far-infrared and sub-mm
are an especially 
promising possibility. The expected surface density of starbursts
associated with the formation of giant ellipticals  
is roughly $2000 h~ (t_{sb}/(10^8{\rm yr})~ {\rm deg}^{-2}$,
where we have adopted $z=3$ as the burst redshift, and an $\Omega = 1$
cosmology.
The typical bolometric luminosity is a few $\times 10^{13} $ \Lsun\ , 
comparable to that of ultraluminous starburst galaxies and quasars.
The predicted flux at $300 \mu$ is about 10 mJy, given
$z_f = 3$ and the bolometric correction such that
$L_{60\mu}/L_{bol}$ is a few percent, by analogy with 
objects such as IRAS F10214+4724.
For lower formation redshifts, the fluxes of individual sources are
higher, but the surface density is lower. Similarly, the scaling
with the starburst timescale is such that the surface densities
increase and the individual fluxes decrease with increasing $t_{sb}$.
If ellipticals at the current epoch are the result of subsequent,
mostly stellar, mergers of several earlier starbursts, then the
predicted surface densites go up by a factor of several and the 
individual fluxes are reduced by the same factor.

Hence coverage of $\sim 10 $ square arc-minutes with a deep survey 
by ISO in the far-infrared, or at submillimeter wavelengths by a bolometer 
array such as is available at IRAM, CSO or (soon) at JCMT, should lead 
to detections of the dust-shrouded starbursts that, we have suggested, 
herald elliptical formation. 
It is conceivable that quasars are not completely 
unrelated to our proposed protoelliptical starburst phase.
A burst of massive star formation need not preclude the presence of
an AGN (Djorgovski 1994), and some models suggest a connection between
a massive central starburst and the formation of an active nucleus
({\it e.g.} Norman \& Scoville 1988). 
It is an intriguing coincidence that abundance ratios in quasar 
emission line regions are also indicative of a massive star-dominated  
nucleosynthetic yield from Type II supernovae (Hamann \& Ferland 1993,
Matteucci \& Padovani 1993).

	The model presented here also makes predictions for observational
tests of elliptical galaxy evolution. If the $M/L$  
trend of ellipticals is a result of remnants from a burst of massive
star formation, the Fundamental Plane of ellipticals is expected to 
evolve in a way very similar to that of passive evolutionary models, 
at least until the redshift of the burst is approached. In contrast,
models in which the $M/L$ trend reflects age variations would
be expected to show rather rapid changes with redshift in the
slope of the Fundamental Plane. Another prediction of the model is
that $M/L$ residuals should correlate with [Mg/Fe] residuals.
Furthermore, the small scatter about the Fundamental Plane
provides constraints on the evolution of elliptical galaxies.
If the mass fraction in the top-heavy mode of star formation is
determined solely by the mass of the protoelliptical at
the time of the starburst, then variations in the history of
subsequent mergers and accretions will introduce scatter about
the Fundamental Plane for any given final elliptical galaxy mass.
Although dependent on cosmology and the details of the merging
process, it is likely that the mass distribution of
captured objects will be steeply declining, so that the resulting dispersion
need not be excessive.  Given the observed dispersion about the Fundamental 
Plane $\sigma(M/L)_{B} \simlt 20\%$, and that our model attributes
about one-third of the mass of a typical elliptical to
stellar remnants of the merger-induced starburst,
a dispersion of up to a factor of $50\%$ in merged masses is acceptable.

%

	Finally, our model makes predictions for the history of the
enrichment of the intracluster medium. A burst of massive star formation 
during the formation of elliptical galaxies will lead to early enrichment 
of the ICM. Thus observations of X-ray gas in moderate and high redshift 
clusters would be expected to show metallicities which are similar 
to those in galaxy clusters today, or even higher if infall has diluted 
the metallicites of rich clusters at low redshift. This prediction 
is in agreement with the ASCA observation of Abell 370 at $z=0.37$ 
(Bautz {\it et al.} 1994).


\acknowledgments
 
We acknowledge interesting discussions with Keith Ashman, Michel Casse,
George Lake, and Jean-Loup Puget.
We also thank the referee, George Djorgovski, for suggestions which 
improved the presentation of the paper.
S.E.Z. acknowledges support from NASA through grant
number HF-1055.01-93A awarded by the Space Telescope Science Institute,
which is operated by the Association of Universities for Research in
Astronomy, Inc., for NASA under contract NAS5-26555.
The research of J.S. has also been supported in part by grants from 
NASA and NSF.

\clearpage

\centerline {\bf Figure Captions}

\noindent Figure 1 - A plot showing the variation of mass-to-light
ratio with elliptical galaxy luminosity and the fraction of the
total mass in remnants. The upper half of the plot describes the 
mass-to-light ratio, with the top line being the observed M/L vs. L 
relationship, and the straight lines representing various choices 
for the luminosity at which remnants from a phase of high-mass
star formation begin to account for the M/L trend.
The masses of this population of remnants are simply the difference
between the upper curve and the straight lines, and are shown graphically
in the lower half of the plot.

\noindent Figure 2 - A plot of the mass densities of stellar rements
as a function of elliptical galaxy luminosity. This quantity is a 
convolution of the remnant masses as a function of galaxy luminosity 
in Figure 1, with the adopted luminosity function for early-type galaxies.
For reference, the elliptical galaxy luminosity function is plotted as 
a solid line above the others. This plot shows that the dominant contribution
of remnant mass (and thus metals and background light) is
from ellipticals at about 2 $L_{*}$. The integral of these curves gives
the total mass density of remnants for the various models.

\noindent Figure 3 - This plot compares the IMFs considered for
the top-heavy starburst with a standard IMF. The top-heavy IMFs
are given by the dashed lines, and the normal IMF by the solid line.
The various dashed lines correspond to IMFs with a slope of
-1, -1.5, and -2. The normalization between the starburst and
normal population shown is that for an $L_{*}$ elliptical galaxy.

\noindent Figure 4 - This plot shows the predicted Fe mass in the Coma
cluster as a function of the luminosity at which remnants of the
massive stellar population beging to contribute to mass-to-light
ratio. The observed value for the Coma cluster is shown as the solid 
line. The model provides a good fit to the data for the fiducial case 
and $h$ close to 0.5, or for cutoff luminosity closer to that of case 3 
if $h$ is closer to 1.0. Note that this plot overestimates the Fe production
of the model because it does {\it not} account the metals retained by
the ellipticals.

\noindent Figure 5 - A plot of the energy density of the extragalactic 
background light produced by the massive star population, as a function 
of the elliptical galaxy luminosity at which it begins to come into play.
The observed upper limit at 4000 \AA\ is shown as a solid line.
This is the same energy density as the upper limit at $240 \mu$ given
by Hauser (1995). Much like the Fe prediction in Figure 4,
the model provides a good fit to the data for low $h$ and the fiducial
cutoff luminosity, and requires a somewhat higher cutoff luminosity
if $h$ is high.
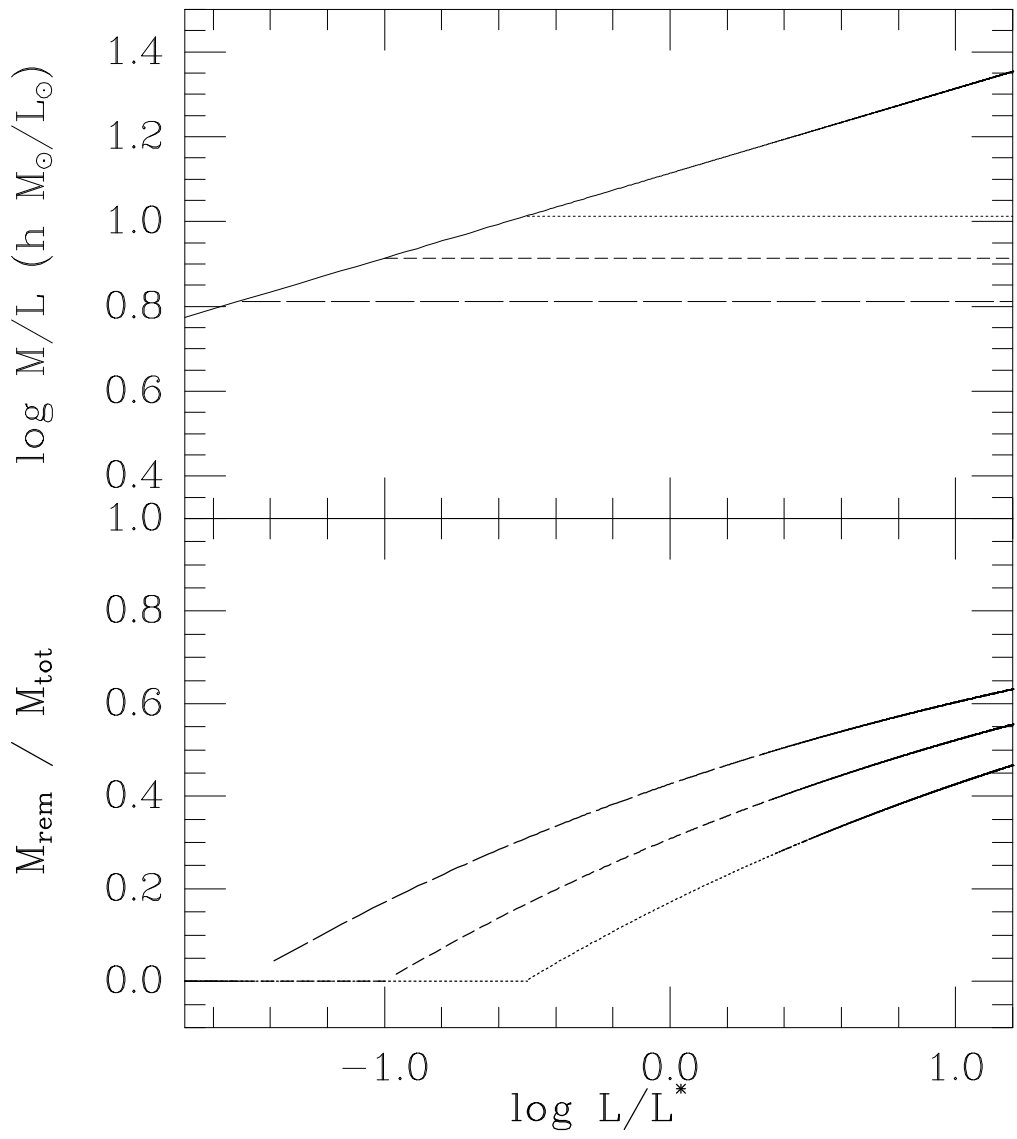
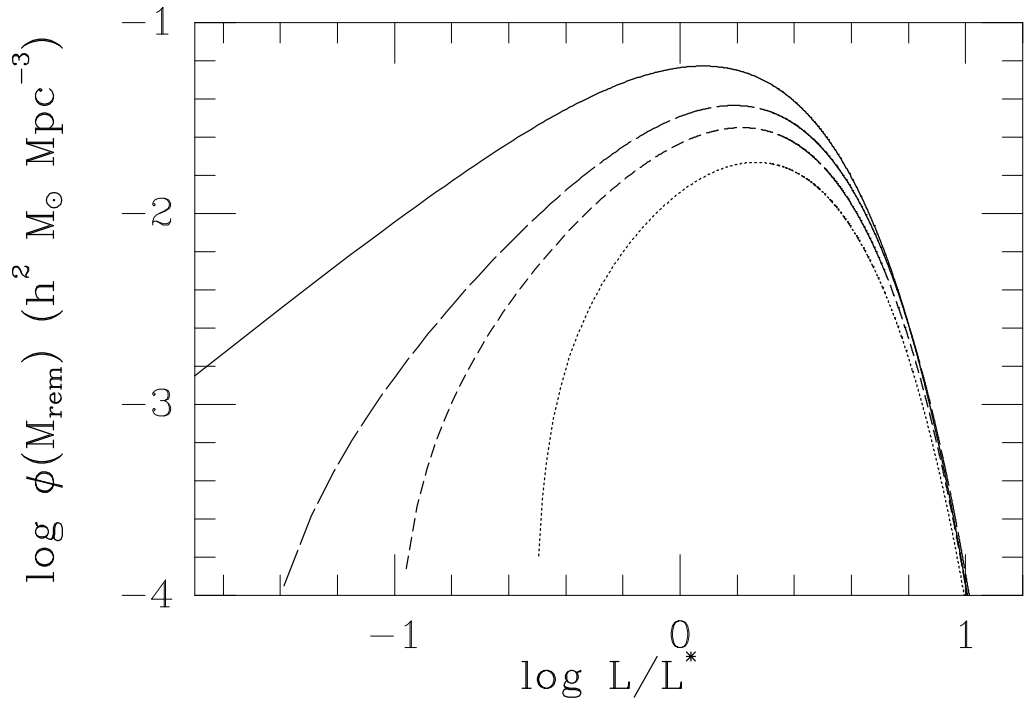
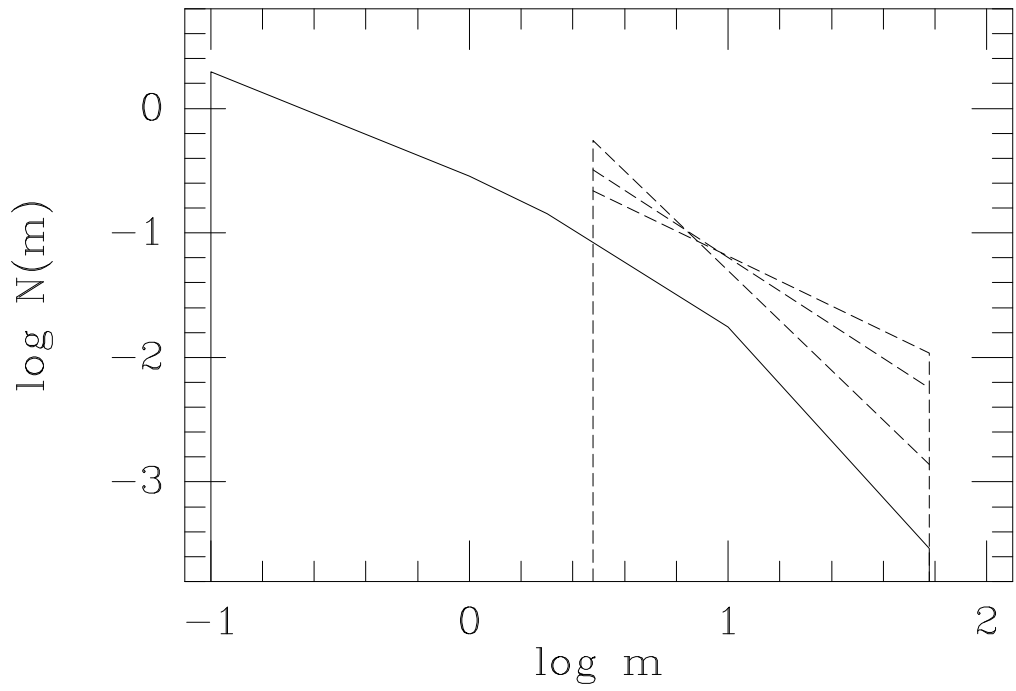
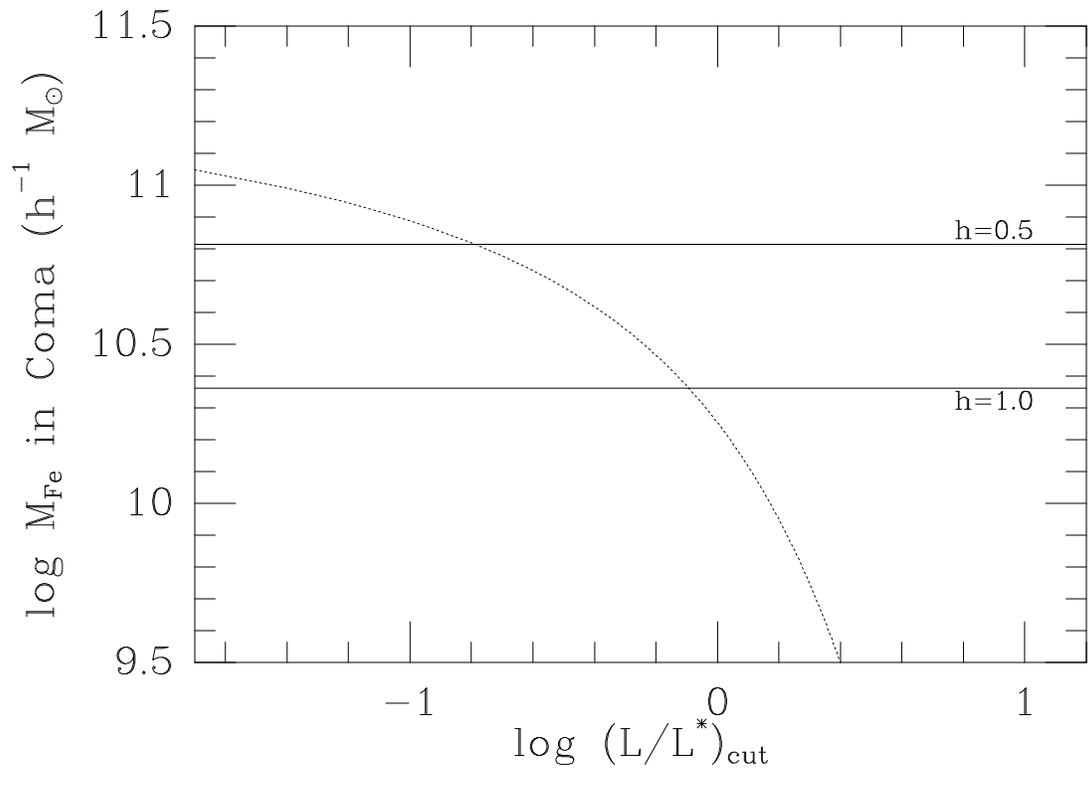
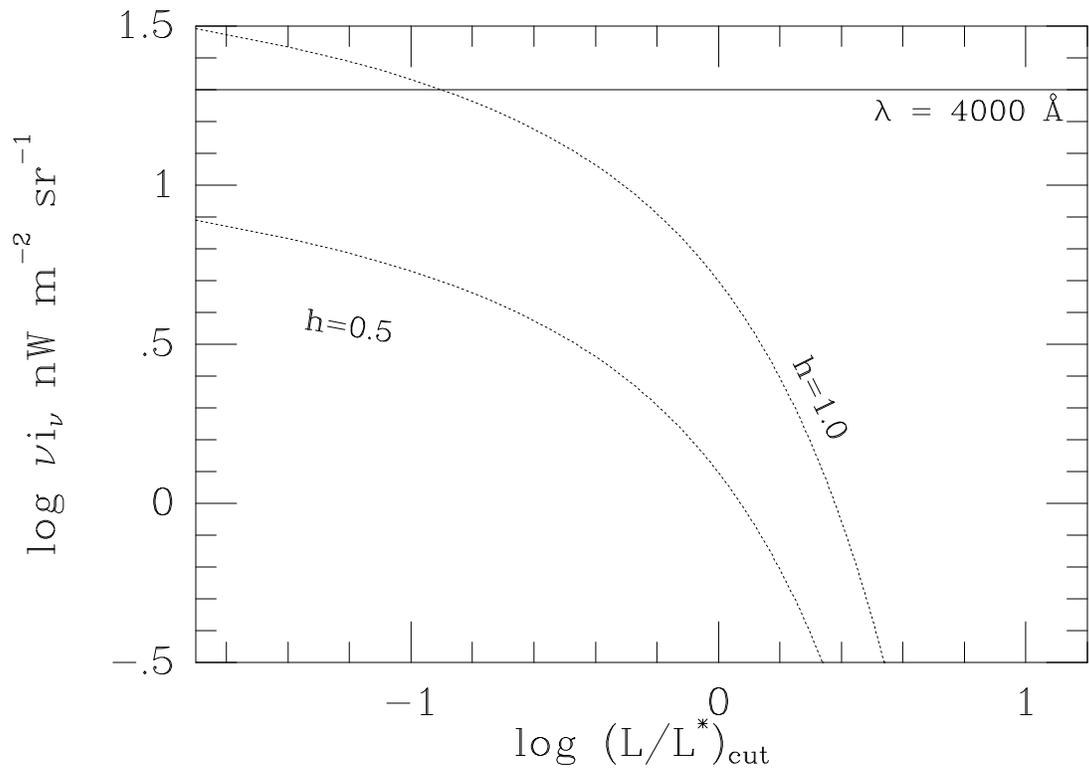
\end{document}